\documentclass[11pt,a4wide]{article}
\pdfoutput=1
\usepackage{jheppub}

\usepackage[english]{babel}
\usepackage[babel]{csquotes}
\usepackage{amsfonts}
\usepackage{array, tabularx}
\usepackage{mathtools}
\usepackage{amsthm}
\usepackage{url}
\usepackage{multicol, multirow}
\usepackage{emptypage}
\usepackage{verbatim}
\usepackage{dsfont}
\usepackage{hyperref}
\usepackage{float}
\usepackage{amsmath}
\usepackage{amssymb,tikz}

\DeclareSymbolFont{bbold}{U}{bbold}{m}{n}
\DeclareSymbolFontAlphabet{\mathbbold}{bbold}

\DeclareMathOperator*{\dDisc}{dDisc}

\theoremstyle{definition}

\theoremstyle{remark}

\theoremstyle{definition}

\allowdisplaybreaks[2]

\usepackage[font=small,format=hang,labelfont={sf,bf}]{caption}

\makeatletter
\DeclareFontFamily{OMX}{MnSymbolE}{}
\DeclareSymbolFont{MnLargeSymbols}{OMX}{MnSymbolE}{m}{n}
\SetSymbolFont{MnLargeSymbols}{bold}{OMX}{MnSymbolE}{b}{n}
\DeclareFontShape{OMX}{MnSymbolE}{m}{n}{
    <-6>  MnSymbolE5
   <6-7>  MnSymbolE6
   <7-8>  MnSymbolE7
   <8-9>  MnSymbolE8
   <9-10> MnSymbolE9
  <10-12> MnSymbolE10
  <12->   MnSymbolE12
}{}
\DeclareFontShape{OMX}{MnSymbolE}{b}{n}{
    <-6>  MnSymbolE-Bold5
   <6-7>  MnSymbolE-Bold6
   <7-8>  MnSymbolE-Bold7
   <8-9>  MnSymbolE-Bold8
   <9-10> MnSymbolE-Bold9
  <10-12> MnSymbolE-Bold10
  <12->   MnSymbolE-Bold12
}{}

\let\llangle\@undefined
\let\rrangle\@undefined
\DeclareMathDelimiter{\llangle}{\mathopen}%
                     {MnLargeSymbols}{'164}{MnLargeSymbols}{'164}
\DeclareMathDelimiter{\rrangle}{\mathclose}%
                     {MnLargeSymbols}{'171}{MnLargeSymbols}{'171}
\makeatother

\title{Taming the $\epsilon$-expansion with \\ Large Spin Perturbation Theory}
\author{Luis F. Alday,}
\author{Johan Henriksson \&}
\author{Mark van Loon}
\affiliation{Mathematical Institute, University of Oxford, Andrew Wiles Building, Radcliffe Observatory Quarter, Woodstock Road, Oxford, OX2 6GG, UK}

\emailAdd{luis.alday@maths.ox.ac.uk}
\emailAdd{johan.henriksson@maths.ox.ac.uk}
\emailAdd{mark.vanloon@maths.ox.ac.uk}

\abstract{We apply analytic bootstrap techniques to the four-point correlator of fundamental fields in the Wilson-Fisher model. In an $\epsilon$-expansion crossing symmetry fixes the double discontinuity of the correlator in terms of CFT data at lower orders. Large spin perturbation theory, or equivalently the recently proposed Froissart-Gribov inversion integral, then allows one to reconstruct the CFT data of intermediate operators of any spin. We use this method to compute the anomalous dimensions and OPE coefficients of leading twist operators. To cubic order in $\epsilon$ the double discontinuity arises solely from the identity operator and the scalar bilinear operator, making the computation straightforward. At higher orders the double discontinuity receives contributions from infinite towers of higher spin operators. At fourth order, the structure of perturbation theory leads to a proposal in terms of functions of certain degree of transcendentality, which can then be fixed by symmetries. This leads to the full determination of the CFT data for leading twist operators to fourth order.  
}

\begin{document}
\maketitle

%--------------------------------------------------------------------------------------

\section{Introduction}
Conformal field theories (CFTs) are characterised by a set of local primary operators, labelled by their scaling dimension $\Delta$ and Lorentz spin $\ell$. The set of all $\{\Delta,\ell\}$ is denoted the spectrum of the theory. An important property of CFT is that these operators satisfy an algebra, the operator product expansion (OPE). The structure constants of this algebra, denoted OPE coefficients, together with the spectrum constitute the {\it CFT data}. Using the OPE, any correlator of local operators can be decomposed into conformal blocks and expressed in terms of the CFT data. Associativity of the OPE is equivalent to the statement that decompositions along different channels should lead to the same final answer. This leads to strong constraints on the CFT data, especially when supplemented with other physical requirements, such as unitarity \cite{Rattazzi:2008pe}. 

In the Lorentzian regime correlators develop singularities when two operators become null separated. The basic realisation of the analytic bootstrap is that singularities in one channel are a consequence of high spin operators being exchanged in the dual channel. This idea was first used in \cite{Fitzpatrick:2012yx,Komargodski:2012ek} to constrain the large spin sector of generic CFTs in the non-perturbative regime. The perturbative regime was subsequently analysed in \cite{Alday:2013cwa} for conformal gauge theories. From these developments the results of \cite{Alday:2007mf} for large spin operators could be understood as arising from crossing symmetry. In a series of papers \cite{Alday:2015eya,Alday:2015ewa} an algebraic machinery to compute the CFT data as a series in inverse powers of the spin was developed. This led to the proposal of a large spin perturbation theory \cite{Alday:2016njk}, a method to compute the CFT data to all orders in inverse powers of the spin from the singularities in the dual channel. The essence of these developments is that the full CFT data can be reconstructed from the singularities of the correlator (to be precisely defined below), up to ambiguities for finite, usually low, values of the spin. A drawback of the method is that these ambiguities are generally not under control. In some perturbative contexts, such as in the weakly coupled CFTs studied in \cite{Alday:2016jfr}, crossing symmetry itself constrains the ambiguities and the CFT data can be extrapolated down to spin two. In a remarkable paper \cite{Simmons-Duffin:2016wlq} it was shown that this is the case even in a non-perturbative context.\footnote{See \cite{Cornagliotto:2017snu} for another very interesting application to non-perturbative theories and \cite{Qiao:2017xif} for a discussion on tauberian theorems that justify these extrapolations.} All this was put on firmer ground in a beautiful paper \cite{Caron-Huot:2017vep} which proved that indeed the CFT data is an analytic function of the spin and arises solely from the singularities of the correlator. This was done through an inversion formula analogous to the Froissart-Gribov integral. Requiring the correct Regge behaviour for the CFT correlator also precludes all ambiguities for spin higher than one. The inversion formula of \cite{Caron-Huot:2017vep} can not only be regarded as the resummed version of large spin perturbation theory, but it also proves that its results indeed do resum into analytic functions of the spin. 

In this paper we will apply the method of large spin perturbation theory to the Wilson-Fisher (WF) model in $d=4-\epsilon$ dimensions. In \cite{Alday:2016jfr} results were obtained for the anomalous dimensions of weakly broken currents to the first non-trivial order in $\epsilon$. Other interesting analytic approaches in the spirit of the conformal bootstrap, that have led to results for the WF model in the $\epsilon$-expansion include \cite{Rychkov:2015naa,Basu:2015gpa,Gliozzi:2016ysv,Roumpedakis:2016qcg,  Liendo:2017wsn,Gliozzi:2017gzh}. In a series of papers \cite{Gopakumar:2016wkt,Gopakumar:2016cpb, Dey:2017fab,Dey:2016mcs} a proposal has been put forward for an alternative method to compute CFT data analytically. In this approach one uses Mellin space and crossing symmetry is built in. Consistency with the OPE then constrains the CFT data. This method has been applied to the WF model in the $\epsilon$-expansion leading to impressive results. More precisely, the CFT data for weakly broken currents has been obtained to cubic order in $\epsilon$. The purpose of the present paper is first to show how these results can be recovered from the perspective of large spin perturbation theory or equivalently from the inversion integral mentioned above. To cubic order the relevant divergences of the correlator arise, via crossing symmetry, from just two operators in the crossed channel: the identity operator and the bilinear scalar operator. This makes our derivation very simple: in the present framework it essentially involves a first-order computation. The simplicity of our method is also manifest when dealing with the $O(N)$ model where the results to cubic order follow straightforwardly from those for $N=1$. A remarkable feature of our computation is that the convergence properties of the inversion integral allow to extrapolate the results down to spin zero. Conservation of the stress tensor together with a matching condition for spin zero lead to two non-trivial constraints, that allow to fix not only the dimension of the external operator but also the dimension of the scalar operator $\varphi^2$. We then move on to the computation at fourth order. In this case the divergences of the correlator are more involved and arise from infinite towers of operators with arbitrarily large spins. The computation is complicated by the appearance of new operators in the OPE at quadratic order. A remarkable feature of these operators, together with intuition from perturbation theory, makes it possible to guess their contribution to the divergence, and hence to determine the CFT data of weakly broken currents to fourth order. The results for the anomalous dimensions agree with those in the literature, computed by Feynman techniques, while the OPE coefficients are a new result. From the latter we deduce the central charge of the WF model to fourth order in the $\epsilon$-expansion:
\begin{equation}
\frac{C_T}{C_{\text{free}}}= 1 - \frac{5}{324} \epsilon^2 - \frac{233}{8748}\epsilon^3 - \left(\frac{100651}{3779136}-\frac{55}{2916}\zeta_3\right) \epsilon^4+\cdots,
\end{equation}
 a new result as well. This paper is organised as follows. The computation up to cubic order is presented in section \ref{thirdorder}. After introducing the basic ingredients we show how to understand the inversion formula from the perspective of large spin perturbation theory. Since we are dealing with leading twist operators, the inversion problem for $SL(2,\mathbb{R})$ suffices. Then we proceed to obtain the CFT data for leading twist operators, up to this order, from the double discontinuity of the correlator. We also show how to generalise these results to the $O(N)$ model. In section \ref{fourthorder} we tackle the problem to fourth order and give the full answer for the anomalous dimensions and OPE coefficients of leading twist operators. We finish with some conclusions. Appendix \ref{integrals} contains a database of the necessary inversion integrals to compute the CFT data at hand, while appendix \ref{ddisc} contains expressions for double discontinuities at fourth order.   

\section{Lorentzian OPE Inversion in the $\epsilon$-expansion}
\label{thirdorder}
\subsection{Generalities}

Consider the four-point correlator of a scalar field $\varphi$ in a $d$-dimensional CFT
\begin{equation}
\langle \varphi(x_1) \varphi(x_2)  \varphi(x_3)  \varphi(x_4) \rangle = \frac{{\cal G}(z,\bar z)}{x_{12}^{2\Delta_\varphi}x_{34}^{2\Delta_\varphi}},
\end{equation}
where we have introduced the cross-ratios
\begin{equation}
z \bar z = \frac{x_{12}^2 x_{34}^2}{x_{13}^2 x_{34}^2},~~~~(1-z)(1-\bar z) =  \frac{x_{14}^2 x_{23}^2}{x_{13}^2 x_{34}^2}.
\end{equation}
Crossing symmetry implies
\begin{equation}
\left(\frac{1-z}{z} \right)^{\Delta_\varphi} {\cal G}(z,\bar z)  = \left(\frac{\bar z}{1-\bar z} \right)^{\Delta_\varphi} {\cal G}(1-\bar z,1-z).
\end{equation}
The correlator admits a decomposition in conformal blocks. The s-channel decomposition reads
\begin{equation}
{\cal G}(z,\bar z) = \sum_{\Delta,\ell} a_{\Delta,\ell} (z \bar z)^{\tau/2} g_{\Delta,\ell}(z,\bar z),
\end{equation}
where $g_{\Delta,\ell}(z,\bar z)$ are the $d-$dimensional conformal blocks and the twist $\tau=\Delta-\ell$ is the dimension minus the spin. Assume there is a free point where the correlator reduces to that of generalised free fields (GFF)
\begin{equation}
{\cal G}^{(0)}(z,\bar z) = 1+ (z \bar z)^{\Delta_\varphi} + \left(\frac{z \bar z}{(1-z)(1-\bar z)}\right)^{\Delta_\varphi}.
\end{equation}
The intermediate operators are the identity and towers of bilinear operators of twist $2\Delta_\varphi +2n$ and spin $\ell$. We will be interested in leading twist operators with $n=0$. In this case the OPE coefficients read
\begin{equation}
a^{(0)}_{\ell} = \frac{2 \left((\Delta_\varphi )_\ell\right){}^2}{\ell! (\ell+2 \Delta_\varphi -1)_\ell},
\end{equation}
where $(a)_n$ is the Pochhammer symbol. As we show below, these OPE coefficients are fixed by the structure of divergences of the correlator. Next we consider perturbations by a small parameter $g$. This introduces a correction to the scaling dimensions and OPE coefficients of the leading-twist operators
\begin{eqnarray}
\Delta_\ell &=& 2\Delta_\varphi + \ell+ \gamma^{(1)}_\ell g + \cdots\nonumber\\
a_{\ell}  &=& a^{(0)}_{\ell} +a^{(1)}_{\ell} g+ \cdots .
\end{eqnarray}
We will assume that at this order no new operators appear in the OPE $\varphi \times  \varphi$. From the analysis of \cite{Alday:2016njk} it follows that the only solutions consistent with crossing symmetry have finite support in the spin. For generic $\Delta_\varphi$ these solutions can be constructed following \cite{Heemskerk:2009pn}. For the present paper we will be interested in the case $\Delta_\varphi=\frac{d-2}{2}$ at leading order in $g$. In this case it was proven in \cite{Alday:2016jfr} that crossing symmetry admits a non-trivial solution only around $d=4$, with support for spin zero. We {\it define} the coupling constant $g$ as the anomalous dimension of the bilinear operator with spin zero
\begin{eqnarray}
\Delta_0 = 2\Delta_\varphi + g.
\end{eqnarray}
All other quantities will be computed in terms of this coupling constant. In \cite{Alday:2016jfr} it was also shown that $\Delta_\varphi$ can receive corrections only at order $g^2$. Note that the dimensionality of space-time can differ from four by at most something of order $g$, so that $d=4-\epsilon$ with $g \sim \epsilon$. The correction to the OPE coefficients can be found through an extension of the analysis of \cite{Alday:2016jfr}. Again, the corresponding solution has support only for spin zero and one finds $a_{0}  =a^{(0)}_{0}(1- g +\cdots)$. In summary, for spin two and higher the corrections start at order $g^2$
\begin{eqnarray}
\Delta_\ell &=& 2\Delta_\varphi + \ell+  \gamma^{(2)}_\ell g^2 + \cdots,\qquad\ell=2,4,\cdots,\nonumber\\
a_{\ell}  &=& a^{(0)}_{\ell} +  a^{(2)}_{\ell} g^2 + \cdots,\qquad\ell=2,4,\cdots ,
\end{eqnarray}
and the same is true for the external operator
\begin{eqnarray}
\Delta_\varphi= \frac{d-2}{2} + \gamma^{(2)}_{\varphi} g^2 + \cdots.
\end{eqnarray}
We would like to find the corrections consistent with crossing symmetry. Our method relies on the fact that the double-discontinuity (to be defined below) of the correlator contains all the relevant physical information. Let us explain this in more detail.
\subsection{From large spin perturbation theory to an inversion formula}
Consider a basis of $SL(2,\mathbb{R})$ conformal blocks $f_{\Delta,\ell}(\bar z)$. We find it convenient to introduce the following normalisation
\begin{equation}
f_{\Delta,\ell}(\bar z) = r_{\frac{\Delta+\ell}{2}} k_{\frac{\Delta+\ell}{2}}(\bar z),\qquad r_h= \frac{\Gamma(h)^2}{\Gamma(2h)},
\end{equation}
with $k_h(\bar z)=\bar z^h ~_2F_1(h,h,2h,\bar z)$ a standard hypergeometric function. We are interested in solving the following inversion problem: find $\hat a_\ell$ such that
\begin{equation}
\label{SL2R}
\sum_{\substack{\Delta=2\Delta_\varphi+\ell,\\\ell=0,2,\cdots}} \hat a_\ell f_{\Delta,\ell}(\bar z) = G(\bar z),
\end{equation}
for a given $G(\bar z)$ containing an enhanced singularity as $\bar z \to 1$. By enhanced singularity we mean a contribution which becomes power-law divergent upon applying the Casimir operator a finite number of times, and as such it cannot be obtained by a finite number of conformal blocks. This is equivalent to saying that $G(\bar z)$ contains a double discontinuity. For a correlator the double discontinuity is defined as the difference between the Euclidean correlator and its two analytic continuations around $\bar z=1$
\begin{equation}
 \dDisc [G(\bar z)] \equiv G(\bar z) -\frac{1}{2} G^\circlearrowleft(\bar z)-\frac{1}{2} G^\circlearrowright(\bar z).
\end{equation}
An algorithm to find $\hat a_\ell$ as a series in $1/\ell$ to all orders was developed in \cite{Alday:2016njk}. The idea is the following. First recall that the $SL(2,\mathbb{R})$ conformal blocks are eigenfunctions of a quadratic Casimir operator
\begin{equation}
\bar D f_{\Delta,\ell}(\bar z)  = J^2 f_{\Delta,\ell}(\bar z) ,
\end{equation}
where $\bar D=\bar z^2 \bar \partial(1-\bar z)\bar \partial$ and $J^2=\frac{1}{4}(\Delta+\ell)(\Delta+\ell-2)$ is called the conformal spin. We then assume that $\hat a_\ell \equiv \hat a(J)$ admits an expansion in inverse powers of the conformal spin
\begin{equation}
\hat a(J) = \sum_m \frac{a_m}{J^{2m}}
\end{equation}
and define the following family of functions
\begin{equation}
h^{(m)}(\bar z)= \sum_{\substack{\Delta=2\Delta_\varphi+\ell,\\\ell=0,2,\cdots}}  \frac{f_{\Delta,\ell}(\bar z)}{J^{2m}} .
\end{equation}
From the explicit form of the blocks we can compute
\begin{equation}
h^{(0)}(\bar z)= \sum_{\substack{\Delta=2\Delta_\varphi+\ell,\\\ell=0,2,\cdots}} f_{\Delta,\ell}(\bar z) = \frac{\pi}{4} \left(\frac{\bar z}{1-\bar z} \right)^{1/2}  +\, \text{regular},
\end{equation}
where the regular terms do depend on $\Delta_\varphi$ but are not important for us. The sequence of functions $h^{(m)}(\bar z)$ can then be generated by the inverse action of the Casimir 
\begin{equation}
\label{recursion}
\bar D h^{(m+1)}(\bar z) = h^{(m)}(\bar z).
\end{equation}
The inversion problem (\ref{SL2R}) then amounts to decomposing $G(\bar z)$ in the basis of functions $h^{(m)}(\bar z)$. The precise range of $m$ depends on the specific form of  $G(\bar z)$. The recursion (\ref{recursion}) can be used to systematically construct the functions $h^{(m)}(\bar z)$ and hence find the coefficients $a_m$. More specifically, one matches the double-discontinuity on both sides of (\ref{SL2R}). To make contact with the inversion formula of \cite{Caron-Huot:2017vep} assume there exists a family of projectors $K^{(m)}(\bar z)$ such that
\begin{equation}
\int_0^1 d\bar z K^{(m)}(\bar z) \dDisc \left[ h^{(n)}(\bar z) \right]= \delta^{mn}.
\end{equation}
Having the projectors $K^{(m)}(\bar z)$ we can write
\begin{equation}
\hat a(J)= \int_0^1 d\bar z K(\bar z,J) \dDisc \left[G(\bar z)\right],
\end{equation}
where 
\begin{equation}
K(\bar z,J) = \sum_m \frac{K^{(m)}(\bar z)}{J^{2m}}.
\end{equation}
As will be clear momentarily, the precise form of these projectors will not be necessary. Acting on both sides of (\ref{SL2R}) with the Casimir operator $\bar D$ and integrating by parts we obtain
\begin{equation}
\left( \bar D^\dagger-J^2\right) K(\bar z,J) =0
\end{equation}
where we have assumed the absence of boundary terms and $\bar D^\dagger = \bar \partial (1-\bar z) \bar \partial \bar z^2$. Introducing the notation $J^2=\bar h(\bar h-1)$ we find two independent solutions related by $\bar h \leftrightarrow 1-\bar h$. We will be interested in the one regular for positive $\bar h$. Requiring the inversion formula to give $\hat a(J)=1$ for $G(\bar z)=h^{(0)}(\bar z)$ fixes the overall normalisation. We find it convenient to use an integral representation that leads to the following result
\begin{equation}
\hat a(\bar h)= \frac{2 \bar h-1}{\pi^2} \int_0^1 dt d\bar z \frac{\bar z^{\bar h-2}(t(1-t))^{\bar h-1}}{(1-t \bar z)^{\bar h}}  \dDisc \left[G(\bar z)\right].
\end{equation}
Integrating over $t$ leads to the inversion formula presented in \cite{Caron-Huot:2017vep}. For all the inversions needed in this paper it will be convenient to integrate first over $\bar z$.

While this discussion is not a rigorous derivation of the inversion formula, for a derivation see \cite{Simmons-Duffin:2017nub}, it explains its relation to large spin perturbation theory. In appendix \ref{integrals} we give several results relevant for our computations below. In all cases the integral is convergent in the region $\bar h>1$. For our application below this means the integral converges and is expected to give the right answer for $\ell > 0$. Below we will discuss the case $\ell=0$ in more detail. 

\subsection{Inverting discontinuities in the $\epsilon$-expansion}
Let us return to the correlator introduced at the beginning of this section. We will use the inversion formula to compute the CFT data of leading twist operators in an expansion to cubic order in $\epsilon$ (or rather $g$). Crossing symmetry implies
\begin{equation}
\label{crossinglt}
\sum_{\substack{\Delta=\tau_\ell+\ell,\\\ell=0,2,\cdots}} \hat a_\ell z^{\tau_\ell/2} f_{\Delta,\ell}(\bar z) = z^{\Delta_\varphi} \left. \left( \frac{\bar z}{1-\bar z}\right)^{\Delta_\varphi} {\cal G}(1-\bar z,1-z)\right|_{\text{small $z$}},
\end{equation}
where the sum runs over leading twist operators with $\tau_\ell=2\Delta_\varphi+g^2 \gamma_\ell^{(2)}+\cdots$ and the OPE coefficients are related to $\hat a_\ell$ by $a_\ell= \hat a_\ell r_{\frac{\tau_\ell}{2}+\ell}$. According to our discussion above, the CFT data appearing on the l.h.s. of (\ref{crossinglt}) can be recovered from the double-discontinuities of the r.h.s. Up to cubic order in $g$ those are straightforward to compute, as they only arise from the identity operator and the bilinear operator of spin zero, so that
\begin{equation}
\sum_{\substack{\Delta=\tau_\ell+\ell,\\\ell=0,2,\cdots}}\!\! \hat a_\ell z^{\tau_\ell/2} f_{\Delta,\ell}(\bar z) = z^{\Delta_\varphi}\! \!\left. \left(\! \frac{\bar z}{1-\bar z}\!\right)^{\Delta_\varphi}\!\! \left(1 + a_0 (1-\bar z)^{\Delta_0/2} g_{\Delta_0,0}(1-\bar z,1-z) + \text{regular}\right)\right|_{\text{small $z$}}\!\!,
\end{equation}
where, recall, $\Delta_0=2\Delta_\varphi+g$. The regular terms do not contribute to the double-discontinuity to the order we are considering. The $d$-dimensional conformal block for a scalar exchange between two identical scalar operators was given in \cite{Dolan:2000ut} 
\begin{equation}
g_{\Delta,0}(1-\bar z,1-z) = \sum_{m,n=0} \frac{\left(\Delta/2\right)^2_m\left(\Delta/2\right)^2_{m+n}}{m! n! \left( \Delta+1-d/2\right)_m \left(\Delta\right)_{2m+n}} (1-z)^m(1-\bar z)^m(1-z \bar z)^n.
\end{equation}
Note that in order to extract the small $z$ dependence the sum over $n$ has to be performed. Expanding the r.h.s. of (\ref{crossinglt}) in powers of $g$ up to cubic order and keeping only terms that contribute to the double discontinuity we obtain
\begin{align}
\sum_{\substack{\Delta=\tau_\ell+\ell,\\\ell=0,2,\cdots}} \hat a_\ell &z^{\tau_\ell/2} f_{\Delta,\ell}(\bar z) =z^{\Delta_\varphi}  \left( \frac{\bar z}{1-\bar z}\right)^{\Delta_\varphi} + \\
&+ z^{\Delta_\varphi} \bar z^{\Delta_\varphi} a_0 \left( \frac{g^2}{8} \log^2(1-\bar z) \left(1+ \epsilon \partial_\epsilon+g \partial_\Delta \right) + \frac{g^3}{48} \log^3(1-\bar z)\right) g^{(4d)}_{2,0}(1-\bar z,1-z), \nonumber
\end{align}
where
\begin{align}
 g^{(4d)}_{2,0}(1-\bar z,1-z) &= \frac{\log \bar z - \log z}{\bar z}, \nonumber\\
 \partial_\epsilon g^{(4d)}_{2,0}(1-\bar z,1-z) &= \frac{(\log \bar z-\log z)(\log \bar z-2)+2\zeta_2}{2 \bar z},  \\
  \partial_\Delta g^{(4d)}_{2,0}(1-\bar z,1-z) &= \frac{\text{Li}_2(1-\bar z) + \log \bar z-\log z - \zeta_2}{\bar z}, \nonumber
\end{align}
and only the small $z$ limit has been considered. We would like to recover the CFT data for leading twist operators from these singularities. This data admits the following decomposition
\begin{align}
\hat a_\ell = \hat a^{(0)}_\ell + g^2 \hat a^{(2)}_\ell+\cdots,\nonumber\\
\tau_\ell = 2\Delta_\varphi + g^2 \gamma^{(2)}_\ell+\cdots, 
\end{align}
where 
\begin{equation}
\hat a^{(0)}_\ell= \frac{2 (2 \bar h-1) \Gamma \left(\bar h+\Delta _{\varphi }-1\right)}{\Gamma \left(\Delta _{\varphi }\right){}^2 \Gamma \left(\bar h-\Delta _{\varphi }+1\right)},\qquad\bar h = \ell+\Delta_\varphi.
\end{equation}
In order to apply the inversion procedure we follow \cite{Alday:2017vkk} and introduce
\begin{eqnarray}
 \hat a_\ell  &=&U^{(0)}_{\bar h} + \frac{1}{2} \partial_{\bar h}  U^{(1)}_{\bar h}, \nonumber\\
 \hat a_\ell  \gamma_\ell &=& U^{(1)}_{\bar h},  
\end{eqnarray}
where we have made clear that the natural variable in which to express $U^{(0)}_{\bar h},U^{(1)}_{\bar h}$ is $\bar h = \ell+ \Delta_{\varphi}$ as opposed to $\ell$. These combinations are the ones that preserve the reciprocity principle proven in \cite{Alday:2015eya}.\footnote{For the present computation we find it convenient to work with this 'bare' $\bar h$ as opposed to the full one, given by $\bar h_f =\frac{\Delta_{\ell}+\ell}{2}$. The standard reciprocity principle for the CFT-data is usually expressed in terms of the full conformal spin $\bar h_f(\bar h_f-1)$. Note that $\bar h_f $ and $\bar h$ coincide to leading order.}
\begin{eqnarray}
U^{(0)}_{\bar h} =   \sum \frac{u^{(0)}_m}{J^{2m}},\qquad U^{(1)}_{\bar h} =   \sum \frac{u^{(1)}_m}{J^{2m}}.
\end{eqnarray}
where in principle these expansions could contain both even and odd powers of $1/J$ as well as logarithmic insertions. In terms of these expansions we obtain
\begin{align}
\sum_m z^{\Delta_\varphi} &\left( u_m^{(0)} + \frac{1}{2} \log z u_m^{(1)} \right)h^{(m)}(\bar z) =z^{\Delta_\varphi}  \left( \frac{\bar z}{1-\bar z}\right)^{\Delta_\varphi} + \\
&+ z^{\Delta_\varphi} \bar z^{\Delta_\varphi} a_0 \left( \frac{g^2}{8} \log^2(1-\bar z) \left(1+ \epsilon \partial_\epsilon+g \partial_\Delta \right) + \frac{g^3}{48} \log^3(1-\bar z)\right) g^{(4d)}_{2,0}(1-\bar z,1-z). \nonumber
\end{align}
This has exactly the form of the inversion problem discussed above. With the inversion formulas given in appendix \ref{integrals} we find
\begin{align}
U^{(0)}_{\bar h} =&\hat a^{(0)}_\ell+(2 \bar h-1)\left( -\frac{ g^2}{(\bar h-1)^2 \bar h^2}   + \frac{ \zeta _2 (\bar h-1) \bar h+1}{(\bar h-1)^2 \bar h^2} g^2 \epsilon -\frac{\zeta _2 (\bar h-1) \bar h-S_1}{(\bar h-1)^2 \bar h^2} g^3\right) + \cdots, \nonumber \\
U^{(1)}_{\bar h} =& \frac{2(1-2 \bar h)}{(\bar h-1) \bar h} g^2 +\frac{2 (2 \bar h-1)}{(\bar h-1) \bar h} g^2 \epsilon + \frac{2 (2 \bar h-1) S_1}{(\bar h-1) \bar h} g^3+\cdots,
\end{align}
where $S_{1}$ denotes the harmonic number with argument $\bar h-1$. These results encode the full CFT data for leading twist operators to cubic order. They translate easily into the standard anomalous dimensions and OPE coefficients and agree exactly with those obtained previously in \cite{Gopakumar:2016cpb}. 

\subsection{Matching conditions at low spin}

\begin{figure}
\centering
\setlength{\unitlength}{1mm}
\begin{picture}(100,63)
\put(0,0){\includegraphics[width=100mm]{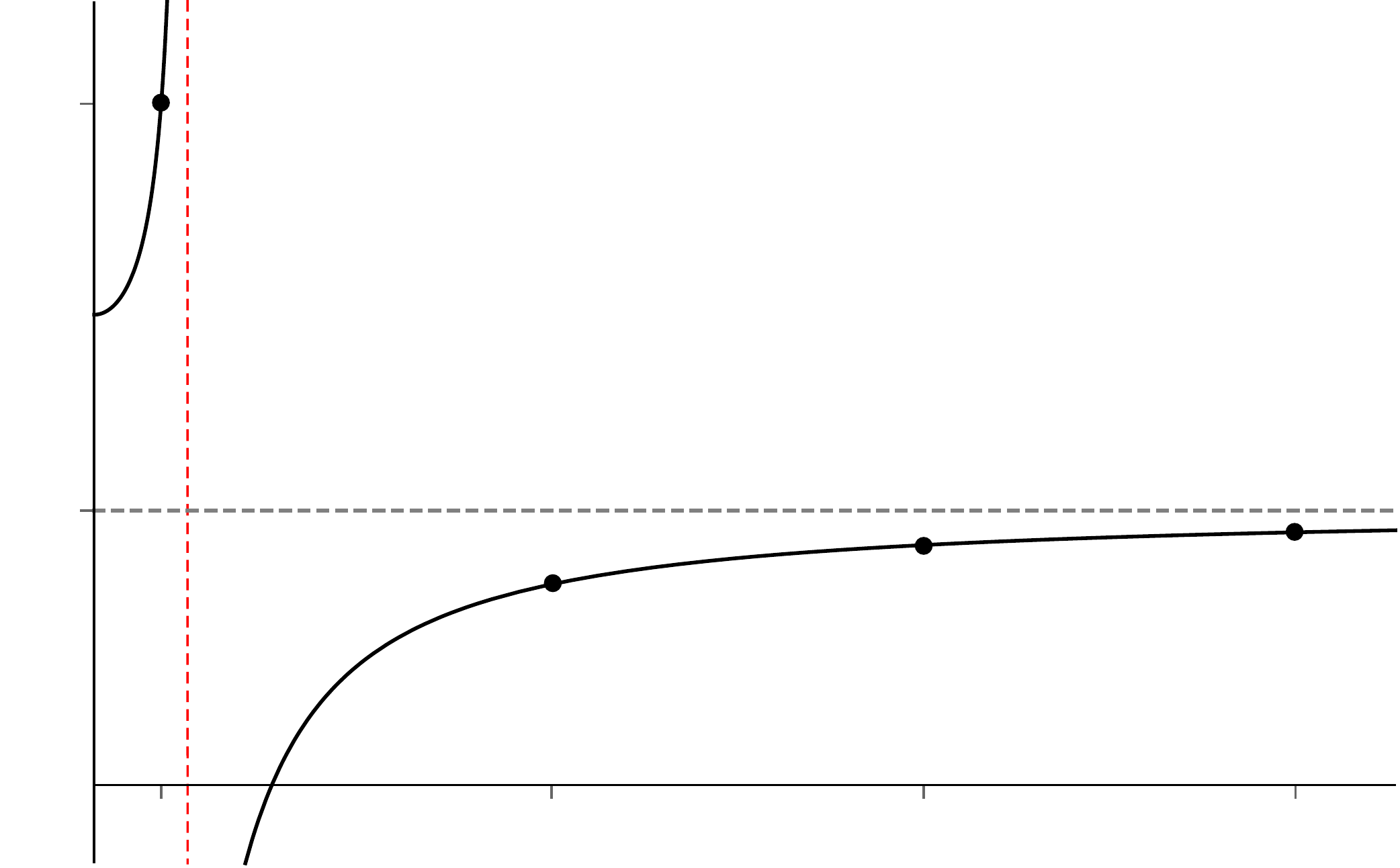}}
\put(10.6,1){$0$}
\put(38.3,1){$2$}
\put(65.1,1){$4$}
\put(91.5,1){$6$}
\put(99,7){$\ell$}
\put(3,61){$\tau_\ell$}
\put(-1,53){$\Delta_0$}
\put(-2,25){$2\Delta_\varphi$}
\end{picture}
\caption{
Schematic graph of $\tau_\ell$. As we move from spin two to spin zero we move to the left of the pole at $\bar h_f=1$, denoted by a red line. Note the change of sign in the correction. Assuming the standard continuation in (\ref{deltaell}), we reproduce the correct dimension on both sides. 
}\label{fig:plot}
\end{figure}
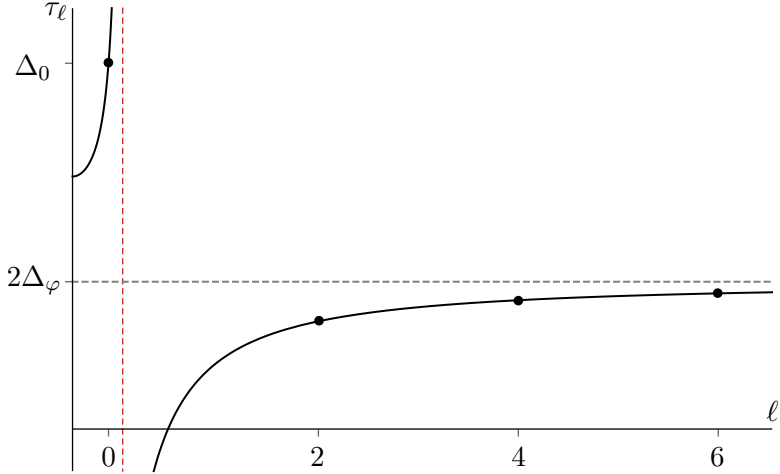

Let us write the result we have just obtained for the anomalous dimensions in terms of the full $\bar h_f$, defined as $\bar h_f = \ell+ \Delta_\varphi+ \frac{1}{2} \gamma_{\ell}$. We obtain
\begin{equation}
\label{deltaell}
\Delta_\ell =2 \Delta_\varphi+ \ell - \frac{g^2}{(\bar h_f-1) \bar h_f} + \frac{g^2 \epsilon+(g^3-g^2\epsilon)S_1}{(\bar h_f-1) \bar h_f} + \cdots
\end{equation}
These results followed only from crossing symmetry of a single correlator and the inversion procedure used in this work shows that they basically follow from a one-loop computation (since squares of anomalous dimensions will generate discontinuities only at quartic order). 
We now impose two further matching conditions at low values of the spin
\begin{eqnarray}
\Delta_2&=&d,\\
\Delta_0 &=& 2\Delta_\varphi+g.
\end{eqnarray}
The first condition is implied by the existence of a conserved stress tensor and fixes the dimension of the external operator
 \begin{equation}
 \Delta_\varphi = 1-\frac{1}{2}\epsilon + \frac{1}{12} g^2 -\frac{1}{8} g^3 + \frac{11}{144} g^2 \epsilon + \cdots.
 \end{equation}
The second condition arises from the requirement that the inversion results can be extrapolated down to spin zero\footnote{We would like to thank Aninda Sinha for suggesting this idea.}. For $\epsilon,g \neq 0$, in order to reach $\ell=0$ we need to continue $\Delta_\ell$ to the left of the pole at $\bar h_f=1$. We will assume the standard continuation across a pole, {\it i.e.} the expression (\ref{deltaell}) remains valid also in this region. This is summarised in Figure~\ref{fig:plot}. 
Note that in the epsilon expansion $\bar h_f-1 \sim \epsilon$, so that the limit is somewhat subtle. To leading order we obtain the following relation
 \begin{equation}
-g \epsilon + 3 g^2=0.
 \end{equation}
This equation has two solutions. One corresponds to the free theory with $g=0$ and the other corresponds to
 \begin{equation}
g = \frac{1}{3}\epsilon + \cdots,
\end{equation}
fixing the relation between $g$ and $\epsilon$. Plugging this into the expression for $\Delta_\varphi$ we obtain
 \begin{equation}
 \Delta_\varphi = 1-\frac{1}{2}\epsilon + \frac{1}{108} \epsilon^2+ \cdots,
 \end{equation}
which exactly agrees with the well known value for the WF model! The order $g^4$ results obtained in the next section allow us to go one order further, and find the relation
 \begin{equation}
g = \frac{1}{3}\epsilon + \frac{8}{81}\epsilon^2+ \cdots.
\end{equation}
Fixing the relation between $g$ and $\epsilon$, and all the quantities entering the problem. 

\subsection{$O(N)$ model}
The method used in this paper generalises to the $O(N)$ model immediately. Let us consider the WF model with $N$ scalar fields $\varphi_i$ with global $O(N)$ symmetry in $d=4-\epsilon$ dimensions. We can now consider the four point correlator of the fundamental field $\varphi_i$. Intermediate operators decompose into the singlet (S), symmetric traceless (T) and anti-symmetric (A) representations of $O(N)$. It is convenient to write the crossing equations as
\begin{eqnarray}
\label{crossON}
f_S(z,\bar z) &=& \frac{1}{N}f_S(1-\bar z,1-z) + \frac{N^2+N-2}{2N^2} f_T(1-\bar z,1-z) +\frac{1-N}{2N} f_A(1-\bar z,1-z),\nonumber\\
f_T(z,\bar z) &=&f_S(1-\bar z,1-z)+ \frac{N-2}{2N} f_T(1-\bar z,1-z) +\frac{1}{2} f_A(1-\bar z,1-z),\\
f_A(z,\bar z) &=&-f_S(1-\bar z,1-z) + \frac{2+N}{2N} f_T(1-\bar z,1-z) +\frac{1}{2} f_A(1-\bar z,1-z), \nonumber
\end{eqnarray}
where $f_{R}(z,\bar z) =((1-z)(1-\bar z))^{\Delta_\varphi}  {\cal G}_{R}(z,\bar z)$. The crossing equations at leading order have been analysed in \cite{Alday:2016jfr} with the methods of large spin perturbation theory. Again, at leading order the fundamental field does not acquire any corrections while
\begin{equation}
\gamma^{(1)}_{\varphi^2_S} = g \equiv g_S,~~~\gamma^{(1)}_{\varphi^2_T} = \frac{2}{2+N}g + \cdots \equiv g_T.
\end{equation}
In order to reconstruct the CFT data from double discontinuities we note that these arise from the identity operator, present in the singlet representation, and the bilinear fields in the singlet and symmetric-traceless representation, which acquire an anomalous dimension at order $g$. By looking at the double discontinuity of the identity operator on the r.h.s. of the crossing equations (\ref{crossON}) we see that at leading order the OPE coefficients of the symmetric traceless and anti-symmetric representations are exactly as before, while those of the single representation have an extra factor of $1/N$. 
 \begin{equation}
 \hat a^{(0)}_{A/T,\ell} = \hat a^{(0)}_{\ell},\qquad \hat a^{(0)}_{S,\ell} = \frac{1}{N} \hat a^{(0)}_{\ell},
 \end{equation}
where $\ell$ is even for the symmetric-traceless and singlet representations and odd for the anti-symmetric representation. A careful analysis of the crossing conditions also determines the corrections to order $g$ of the OPE coefficients for the spin zero operators:
\begin{equation}
a_{S/T,0} = a^{(0)}_{S/T,0} (1-g_{S/T} + \cdots).
\end{equation}
By looking at the crossing equations (\ref{crossON}) and comparing them with our computation for the $N=1$ case, it is then straightforward to write down the result for  $U^{(1)}_{\bar h}=\hat a_\ell  \gamma_\ell$ for each representation. We obtain
\begin{align}
U^{(1)}_{S,\bar h} = &\frac{1}{N^2}\left( \frac{(2-4 \bar h)}{(\bar h-1) \bar h} g_S^2 +\frac{2 (2 \bar h-1)}{(\bar h-1) \bar h} g_S^2 \epsilon + \frac{2 (2 \bar h-1) S_1}{(\bar h-1) \bar h} g_S^3\right) \nonumber \\
&+ \frac{N^2+N-2}{2N^2}\left( \frac{(2-4 \bar h)}{(\bar h-1) \bar h} g_T^2 +\frac{2 (2 \bar h-1)}{(\bar h-1) \bar h} g_T^2 \epsilon + \frac{2 (2 \bar h-1) S_1}{(\bar h-1) \bar h} g_T^3\right)  + \cdots, \nonumber\\
U^{(1)}_{T,\bar h} = &\frac{1}{N} \left( \frac{(2-4 \bar h)}{(\bar h-1) \bar h} g_S^2 +\frac{2 (2 \bar h-1)}{(\bar h-1) \bar h} g_S^2 \epsilon + \frac{2 (2 \bar h-1) S_1}{(\bar h-1) \bar h} g_S^3\right)\nonumber  \\
&+ \frac{N-2}{2N}\left( \frac{(2-4 \bar h)}{(\bar h-1) \bar h} g_T^2 +\frac{2 (2 \bar h-1)}{(\bar h-1) \bar h} g_T^2 \epsilon + \frac{2 (2 \bar h-1) S_1}{(\bar h-1) \bar h} g_T^3\right)  + \cdots, \nonumber\\
U^{(1)}_{A,\bar h} = &\frac{1}{N} \left( \frac{(2-4 \bar h)}{(\bar h-1) \bar h} g_S^2 +\frac{2 (2 \bar h-1)}{(\bar h-1) \bar h} g_S^2 \epsilon + \frac{2 (2 \bar h-1) S_1}{(\bar h-1) \bar h} g_S^3\right)\nonumber  \\
&- \frac{2+N}{2N}\left( \frac{(2-4 \bar h)}{(\bar h-1) \bar h} g_T^2 +\frac{2 (2 \bar h-1)}{(\bar h-1) \bar h} g_T^2 \epsilon + \frac{2 (2 \bar h-1) S_1}{(\bar h-1) \bar h} g_T^3\right)  + \cdots ,
\end{align}
where again the harmonic number $S_1$ is evaluated at $\bar h-1$. The OPE coefficients can be obtained in exactly the same way. All the results are in full agreement with those obtained in \cite{Dey:2016mcs,Manashov:2017xtt}.\footnote{Now $g_S=\frac{2+N}{8+N}\epsilon+6\frac{(N+2)(N+3)}{(N+8)^3} \epsilon^2+\cdots$ and $g_T= \frac{2}{8+N}\epsilon+\frac{36+4N-N^2}{(N+8)^3}\epsilon^2+\cdots$.}

\section{Results to fourth order}
\label{fourthorder}
\subsection{New operators at second order}
Before proceeding to solve the crossing constraints to higher order, we would like to make the following crucial observation.  At order $g^2$ new intermediate operators are expected to appear, since the Lagrangian contains a quartic interaction vertex. They are of the schematic form $\varphi^2 \Box^n \partial_{\mu_1} \cdots \partial_{\mu_\ell} \varphi^2$ and have twist $\tau=4+2n$ and spin $\ell$. These operators are expected to acquire an anomalous dimension to order $\epsilon$. Hence, they generate a double discontinuity, proportional to the square of their anomalous dimension, to order $g^4$. Furthermore, these operators are highly degenerate in perturbation theory, so that computing this double discontinuity would require solving a mixing problem. The statement that the CFT data can be reconstructed from the double-discontinuities of the correlator is not restricted to leading twist operators and the method described in this paper can again be used to find the leading OPE coefficients of these operators. The steps are very similar to the ones above, and to second order in $g$ we find
\begin{equation}
\label{twist4ope}
a_{4+2n,\ell}= 
     \begin{cases}
       \frac{\Gamma (\ell+2)^2}{\Gamma (2 \ell+3)} \frac{\ell^2+3 \ell+8}{12 (\ell+1) (\ell+2)} g^2 + \cdots &\quad\text{for $n=0$,}\\ \\
       {\cal O}(g^4) &\quad\text{for $n \neq 0$.}\\     
       \end{cases}
\end{equation}
This is a somewhat surprising result:  only operators with approximate twist four appear at this order.\footnote{As a byproduct, this result justifies an ansatz made in \cite{Liendo:2012hy}, where the vanishing of OPE coefficients involving operators with $n \neq 0$ was assumed. 
} As we will see, this constrains the possible structure of double discontinuities at fourth order and it will allow us to solve the problem completely. Given the convergence of the inversion integrals we expect these results to be valid down to spin zero. 

\subsection{Solving the inversion problem at fourth order}
The contribution arising from leading twists operators in a perturbative $\epsilon$-expansion can be encoded as follows
\begin{align}
\left. {\cal G}(z,\bar z) \right|_{\text{small $z$}} = \sum_m z^{\Delta_\varphi} &\left( u_m^{(0)} + \frac{1}{2} \log z \, u_m^{(1)}+ \frac{1}{8} \log^2 z\,  u_m^{(2)} + \cdots \right)h^{(m)}(\bar z),
\end{align}
where $u_m^{(q)} \sim g^{2q}$ for small $g$. As before the $u_m^{(q)}$ are the coefficients in the large $J$ expansions of $U^{(q)}_{\bar h}$, whose relation to the usual OPE data is
\begin{align}
\label{datafromU}
\hat a_\ell \left( \gamma_\ell  \right)^p = U^{(p)}_{\bar h} + \frac{1}{2} \partial_{\bar h} U^{(p+1)}_{\bar h}+ \frac{1}{8} \partial^2_{\bar h} U^{(p+2)}_{\bar h}+\cdots.
\end{align}
To order $g^4$ the double-discontinuity of the correlator arises from four distinct contributions, so that
\begin{align}
\left. {\cal G}(z,\bar z) \right|_{\text{small $z$}} = z^{\Delta_\varphi} \left( \left( \frac{\bar z}{1-\bar z}\right)^{\Delta_\varphi}+ I_{\varphi^2} +I_{2} + I_{4} + \, \text{regular} \right).
\end{align}
$I_{\varphi^2}$ denotes the contribution from the scalar bilinear operator. To cubic order it was given in the previous section. It is straightforward to compute it to fourth order and the result is given in appendix \ref{ddisc}. $I_{2}$ denotes the contribution arising from leading twist operators of spin two and higher: the square of their anomalous dimension generates a double-discontinuity at fourth order. Since these operators are non-degenerate, this contribution can be readily computed and it is given in  appendix \ref{ddisc}. As already mentioned, a direct computation of $I_{4}$ would require solving a mixing problem, for instance by considering more general correlators.\footnote{The contribution from twist four operators to the anomalous dimension of leading twist operators starts at order $1/\ell^4$, see \cite{Alday:2007mf}, so that the leading terms in a $1/\ell$ expansion can still be computed without its knowledge. This was done in \cite{Dey:2017oim} by applying directly the methods of \cite{Alday:2015ewa} for isolated operators. Since there is an accumulation point at twist two, one should be careful. In principle the correct procedure from the large spin perspective would be to compute the double discontinuity due to the tower of twist-two operators and then compute the anomalous dimensions from there. The procedure of \cite{Dey:2017oim} is justified since the resulting series are convergent.} However note that at fourth order $I_{4}$ involves four-dimensional conformal blocks evaluated at the classical twist four. This implies the following structure
\begin{equation}
I_{4}= \left( \log z g(\bar z) - \log \bar z g(z) \right) \log^2(1-\bar z),
\end{equation}
where $g(\bar z)$ arises from a sum over twist-four operators 
\begin{equation}
\label{gsum}
g(\bar z) = \frac{1}{8}\sum \eta_\ell k_{2+\ell}(1-\bar z)
\end{equation}
for some $\eta_\ell$ equal to the weighted average, over degenerate operators, of the square anomalous dimensions $\eta_\ell =\langle a_{4,\ell} \gamma^2_{4,\ell}\rangle=\sum_i a_{4,\ell,i} \gamma^2_{4,\ell,i}$. As such it is regular around $\bar z=1$. Furthermore, the structure of the OPE to this order implies the following expansion around $z=0$ \footnote{Specifically, note that in equation (3.4), on the l.h.s. any higher powers $\log^k z$ would have to be generated by higher powers $\gamma_{2,\ell}^k$ of anomalous dimensions, which contribute only at order $g^{2k}$ and higher.}
\begin{equation}
\label{garound0}
g(z) = \alpha_0 \log^2 z + \alpha_1 \log z+ \alpha_2 + \cdots.
\end{equation}
We will now discuss how to fix  $U^{(0)}_{\bar h},U^{(1)}_{\bar h},U^{(2)}_{\bar h}$ to quartic order. Before we proceed, note that the term $\log z g(\bar z)$ in $I_4$ will only contribute to $U^{(1)}_{\bar h}$. Hence $U^{(0)}_{\bar h}$ and $U^{(2)}_{\bar h}$ only require minimal information about $g(z)$, namely only its limit as $z \to 0$. As a result, they could be fully determined in terms of $\alpha_0$  and $\alpha_2$, even without any knowledge of twist four operators. We will be able to do even better than this. 

Let us start with $U^{(2)}_{\bar h}$. From the expressions in appendix \ref{ddisc}, it follows that $I_{\varphi^2}$ and $I_2$ do not contribute to $U^{(2)}_{\bar h}$, as they do not contain a $\log^2z$ piece. The whole contribution arises then from $I_{4}$ and is proportional to $-\alpha_0\log \bar z \log^2(1-\bar z)$. From the results in appendix \ref{integrals} this immediately gives
\begin{equation}
U^{(2)}_{\bar h}=-8 \alpha_0 \frac{4(1-2\bar h)}{\bar h^2(1-\bar h)^2} g^4,
\end{equation}
which exactly agrees with $\hat a_\ell(\gamma_\ell)^2$ to order $g^4$ provided $\alpha_0=1/16$. 

To compute $U^{(1)}_{\bar h}$ one needs to know $g(z)$. The full results for double discontinuities up to cubic order as well as the double discontinuities in appendix \ref{ddisc} suggest that perturbative results for the present correlator organise themselves in pure transcendental functions with discontinuities around $\bar z=0,1$. Furthermore, the degree of these functions increases with the perturbative order in a prescribed way \footnote{More precisely, up to this order we will assume the answer can be written as combinations of polylogarithms of $\bar z$ and $1-\bar z$, without rational functions in front, such that the total degree increases linearly with the loop order. This structure is very familiar in other perturbative contexts.}. If this principle holds then we expect $g(z)$ to be given by a linear combination of the following building blocks
\begin{equation}
\{\log^2 \bar z, \text{Li}_2(1-\bar z),\log^3 \bar z, \log \bar z \,\text{Li}_2(1-\bar z) ,\text{Li}_3(1-\bar z),\text{Li}_3\left(\frac{\bar z-1}{\bar z}\right)  \}.
\end{equation}
The fact that $g(\bar z)$ arises from twist four operators in the dual channel, constrains the possibilities. Furthermore, consistency with (\ref{gsum})  and (\ref{garound0}) leads us to the following result
\begin{equation}
\label{gzb}
g(\bar z) =  \frac{1}{16} \log^2 \bar z +\alpha \left(-\frac{1}{6} \log^3 \bar z-\frac{2}{3} \log z \,\text{Li}_2(1-\bar z)+  \text{Li}_3(1-\bar z)+\text{Li}_3\left(\frac{\bar z-1}{\bar z}\right)\right),
\end{equation}
with a single undetermined coefficient.  We would like to stress that this expression can be systematically tested as an expansion around $\bar z = 1$.  Since $k_{2+\ell}(1-\bar z) \sim (1-\bar z)^{2+\ell}$, to any given order in $(1-\bar z)$ only a finite number of operators contributes and the mixing problem is finite. For instance,  twist four operators with spin zero and two are non-degenerate. Again, their anomalous dimensions can be computed from the discontinuities of the correlator at cubic order, exactly as done above for the leading twist operators\footnote{Alternative, one can also use the method of multiplet recombination, \cite{Rychkov:2015naa}, still purely by bootstrap methods.}, although this would be a somewhat tedious computation. Instead, we will use the well known result  given in \cite{Kehrein:1994ff}. In our conventions $\gamma_{4,0}=3 g+\cdots$ and $\gamma_{4,2}=4/3 g+\cdots$. From (\ref{twist4ope}) we can also read off $a_{4,0}=g^2/6+\cdots $ and  $a_{4,2}=g^2/160 + \cdots$. These values are exactly consistent with the expression for $g(\bar z)$ up to fifth order in $(1-\bar z)$ and furthermore fix $\alpha=-3/2$. With this we find
\begin{equation}
g(z) = \frac{1}{16} \log^2 z -\frac{1}{2}\zeta_2 \log z-\frac{3}{2} \zeta_3 + \cdots,\qquad\text{around $z=0$}.
\end{equation}
We have now all the ingredients to compute $U^{(0)}_{\bar h}$ and $U^{(1)}_{\bar h}$ to fourth order. Using the inversion formulae in appendix \ref{integrals} we find:
\begin{align}
U^{(1)}_{\bar h} = &  \frac{2 (1-2 \bar h)}{(\bar h-1) \bar h} g^2
+  \frac{2 (2 \bar h-1) \left(3+S_1\right)}{(\bar h-1)\bar h}g^3
+  \frac{2 \bar{h}-1}{6 \left(\bar{h}-1\right) \bar{h}} \Bigg(
\frac{6}{\left(\bar{h}-1\right)^2 \bar{h}^2}  + \frac{7+48 S_{-2}}{\left(\bar{h}-1\right) \bar{h}} \nonumber \\
&-9 \zeta _2-6 S_1^2-36 S_1-12 S_{-2}-58\Bigg)g^4+ \cdots,\\
U^{(0)}_{\bar h} = &\hat a_\ell^{(0)} 
+ \frac{ (1-2 \bar h)}{(\bar h-1)^2  \bar h^2} g^2
+  \frac{(2 \bar h-1)}{(\bar h-1)  \bar h}\left(\frac{3+S_1}{(\bar{h}-1)\bar{h}} + 2  \zeta_2 \right)g^3
+ \frac{2 \bar{h}-1}{12 \left(\bar{h}-1\right) \bar{h}}\Bigg(\frac{2}{(\bar{h}-1)^2 \bar{h}^2}  \nonumber\\
& -\frac{56+3 \zeta _2+72 \zeta _3+6 S_1^2+36 S_1-12 S_{-2}}{(\bar{h}-1)   \bar{h}} -106\zeta _2+72 \zeta _3-24 \zeta _2 S_1-54 S_3 \Bigg)g^4 + \cdots,
\end{align}
where the argument of all nested sums, defined in appendix \ref{integrals}, is $\bar h-1$. The CFT data can then be recovered from (\ref{datafromU}). In particular
\begin{equation}
\gamma_\ell = \frac{U^{(1)}_{\bar h} + \frac{1}{2} \partial_{\bar h} U^{(2)}_{\bar h} + \cdots}{U^{(0)}_{\bar h} + \frac{1}{2} \partial_{\bar h} U^{(1)}_{\bar h} + \cdots},
\end{equation}
and the result can be seen to exactly agree with that obtained in \cite{Derkachov:1997pf}\footnote{We would like to thank the authors of \cite{Dey:2017oim} for making us aware of a typo in  \cite{Derkachov:1997pf}.}. In order to fix $\Delta_\varphi$ and $g(\epsilon)$ to this order one could proceed exactly as before: $\Delta_\varphi$ follows again from conservation of the stress tensor while $g(\epsilon)$ follows from the matching condition at spin zero. However, the later result to cubic order would require going to higher orders in our computation. Instead, we will take a shortcut and assume the known value of the dimension of the fundamental field $\Delta_\varphi=1-\frac{\epsilon}{2}+\frac{\epsilon^2}{108}+\frac{109}{11664}\epsilon^3+(\frac{7217}{1259712}-\frac{2}{243}\zeta_3)\epsilon^4+\cdots$. This together with the conservation of the stress tensor gives the relation between $g$ and $\epsilon$:
\begin{equation}
g=\frac{\epsilon}{3}+\frac{8}{81}\epsilon^2+\left(\frac{305}{8748}-\frac{4}{27}\zeta_3 \right)\epsilon^3+\cdots.
\end{equation}
Let us stress however, that the first two orders follow completely from our results, without any additional input, and also the next term could be in principle computed in our formalism. The result for the OPE coefficients is completely new. The most interesting quantity that can be extracted from them is the central charge, related to the OPE coefficient for $\ell=2$. In terms of $\epsilon$ we find
\begin{equation}
\frac{C_T}{C_{\text{free}}}= 1 - \frac{5}{324} \epsilon^2 - \frac{233}{8748}\epsilon^3 - \left(\frac{100651}{3779136}-\frac{55}{2916}\zeta_3\right) \epsilon^4+\cdots,
\end{equation}
 where we have stressed the fact that the contribution proportional to $\epsilon^4$ is also negative. The result to cubic order exactly reproduces what was found in \cite{Gopakumar:2016cpb}. The result to fourth order is new. Setting $\epsilon=1$ we see that this new contribution gets us closer to the highly precise numerical result for the 3d Ising model found in \cite{El-Showk:2014dwa,Kos:2016ysd}. 

\section{Conclusions} 
We have used analytic bootstrap techniques to derive the anomalous dimensions and OPE coefficients of bilinear operators in the WF model in $d=4-\epsilon$ dimensions, to fourth order in the $\epsilon$-expansion. To cubic order the computation is essentially straightforward, since the double discontinuity arises solely from the identity operator and the bilinear scalar. This simplicity is also manifest in the results of the $O(N)$-model, which in our framework follow directly from the results for $N=1$. At fourth order the situation is much more interesting, since two towers of high spin operators, of twist two and four respectively, contribute to the discontinuity. The contribution from twist two operators can be readily computed, while the structure of perturbation theory, together with the explicit form of four-dimensional conformal blocks, allows to make a proposal for the double discontinuity due to twist four operators. This proposal can be systematically tested order by order in powers of $(1-\bar z)$, by solving a finite order mixing problem. In this paper we have not proved such a proposal, but we have checked it to high order. With this result, we have found the CFT data to fourth order. Two further constraints, namely conservation of the stress tensor, together with a continuation to spin zero, allowed to fix the anomalous dimensions of both the scalar operator $\varphi^2$ as well as the dimension of the external operator. All within our framework. 

There are several interesting open problems. A remarkable feature of our computation is the apparent analyticity down to spin zero. This allowed us to reproduce constraints analogous to those of a vanishing beta function. It would be interesting to understand the systematics of this to higher orders, and even non-perturbatively. It would also be interesting to understand the structure of double-discontinuities to higher orders in the $\epsilon$-expansion. Up to fourth order we have observed that the functions that appear have pure transcendentality.  It is tantalising to propose that this persists to higher orders. This would greatly simplify the computation of CFT data.  Another interesting problem is the extension of the present methods to other CFTs. Large spin perturbation theory has been successfully applied to several models at leading order, including cubic models in six dimensions and large-$N$ critical models \cite{Alday:2016jfr}, weakly coupled gauge theories \cite{Alday:2016jfr,Henriksson:2017eej} and fermionic CFTs \cite{vanLoon:2017xlq,Charan:2017jyc}. Another interesting family of CFTs are the multicritical models, studied {\it e.g.} in \cite{Codello:2017hhh}. A natural direction would be to extend these results to higher orders.  It would also be interesting to consider analytic constraints arising from mixed correlators. In the present case one could consider correlators of the fundamental field and the bilinear scalar. The crossing constraints for such a system are expected to be stronger than the ones considered in this paper. 

\section*{Acknowledgements} 
We are grateful to A. Bissi for useful discussions. 
This work was supported partially by ERC STG grant 306260. L.F.A. is a Wolfson Royal Society Research Merit Award holder. M.v.L. was also supported by an EPSRC studentship.

\appendix

\section{Some inversion integrals}
\label{integrals}
In the body of the paper we arrived at the following inversion integral
\begin{equation}
\hat a(J)= \frac{2 \bar h-1}{\pi^2} \int_0^1 dt d\bar z \frac{\bar z^{\bar h-2}(t(1-t))^{\bar h-1}}{(1-t \bar z)^{\bar h}}  \dDisc \left[G(\bar z)\right],
\end{equation}
where % recall 
$J^2=\bar h(\bar h-1)$. As an example, used in the body of the paper, consider 
\begin{equation}
\label{ap}
G(\bar z) = \left( \frac{\bar z}{1-\bar z}\right)^p\quad\to\quad \dDisc \left[ G(\bar z) \right] = 2\sin^2(\pi p)\left( \frac{\bar z}{1-\bar z}\right)^p.
\end{equation}
In this case
\begin{equation}
\hat a_p(J)= \frac{2 (2 \bar h-1) \Gamma (\bar h+p-1)}{ \Gamma (p)^2 \Gamma (\bar h-p+1)}.
\end{equation}
This precise inversion problem was also considered in \cite{Simmons-Duffin:2016wlq}, with exactly the same result. Other inversions used in this paper are presented in tables~\ref{tab:logfree}--\ref{tab:logcont}. In these tables we use the nested harmonic sums $S_{\mathbf{a}}=S_{\mathbf{a}}(\bar h-1) $ which for integer arguments take the values
\begin{equation}
S_{a_1,a_2,\ldots}(n)=\sum_{b_1=1}^n\frac{(\mathrm{sgn}\, a_1)^{b_1}}{b_1^{|a_1|}}\sum_{b_2=1}^{b_1}\frac{(\mathrm{sgn}\, a_2)^{b_2}}{b_2^{|a_2|}}\sum_{b_3=1}^{b_2}\frac{(\mathrm{sgn}\, a_3)^{b_3}}{b_3^{|a_3|}}\cdots.
\end{equation}
For non-integer values of $\bar h$ we make the standard analytic continuation from even arguments $n$, see {\it e.g.} \cite{Albino:2009ci}, so that for instance
\begin{equation}
S_{-2}(x)=\frac{1}{4}\left(\psi^{(1)}\left(\tfrac{x+1}{2}\right)-\psi^{(1)}\left(\tfrac{x+2}{2}\right)\right)-\frac{\zeta_2}{2},
\end{equation}
where $\psi^{(1)}(x)$ is the trigamma function.

To evaluate these inversion integrals is non-trivial, but one can proceed as follows. Expanding the function to invert in powers of $\frac{1-\bar z}{\bar z}$ we are led to the integral entering in (\ref{ap}). We then obtain an expression for the inverted function, as a power expansion for large $\bar h$, which can be resummed. The final result can then be checked numerically, for finite values of $\bar h$, to very high precision.

\begin{table}[H]
\begin{center}
\begin{tabular}{l|l}
$ G(\bar{z}) $ & $\hat{a}(J)$\rule[-1.1ex]{0pt}{0pt}  \\\hline
$\log ^2\left(1-\bar{z}\right)$&$\displaystyle \dfrac{4 \left(2 \bar{h}-1\right)}{(\bar h-1)\bar h}$
\rule{0pt}{4.5ex} \rule[-4.5ex]{0pt}{0pt} \\
$\log^3(1-\bar z)$ & $\displaystyle -\dfrac{24 (2 \bar h-1) S_1}{(\bar h-1) \bar h}$
 \rule{0pt}{4.5ex} \rule[-4.5ex]{0pt}{0pt} \\
$\log^4(1-\bar z)$ & $\displaystyle\dfrac{96 \left(2 \bar{h}-1\right)}{(\bar h-1)\bar h}\left(S_1^2- \zeta _2- S_{-2}\right)$
 \rule{0pt}{4.5ex} \rule[-4.5ex]{0pt}{0pt} \\
$ \log ^2\left(1-\bar{z}\right)\text{Li}_2\left(1-\bar{z}\right)$ & $\displaystyle\dfrac{4 \left(2 \bar{h}-1\right)}{(\bar h-1)\bar h}\left( \zeta _2+2 S_{-2}\right)$
 \rule{0pt}{4.5ex} \rule[-4.5ex]{0pt}{0pt} \\
$ \log ^3\left(1-\bar{z}\right)\text{Li}_2\left(1-\bar{z}\right)$ & $\displaystyle\dfrac{24 \left(2 \bar{h}-1\right)}{(\bar h-1)\bar h}\left( \left( S_{-3}-2 S_{-2,1}\right)-3 \left( S_{-3}-2 S_{1,-2}\right)+3 \zeta _2 S_1-2
   S_3\right)$
 \rule{0pt}{4.5ex} \rule[-4.5ex]{0pt}{0pt} \\
$ \log ^2\left(1-\bar{z}\right)\text{Li}_3\left(1-\bar{z}\right)$ & $\displaystyle\dfrac{4 \left(2 \bar{h}-1\right)}{(\bar h-1)\bar h}\left(-2 \left( S_{-3}-2 S_{1,-2}\right)+\zeta _3+2 \zeta _2 S_1-2 S_3\right)$
 \rule{0pt}{4.5ex} \rule[-4.5ex]{0pt}{0pt} \\
$ \log ^2\left(1-\bar{z}\right)\text{Li}_3\left(\dfrac{\bar{z}-1}{\bar{z}}\right)$ & $\displaystyle\dfrac{4 \left(2 \bar{h}-1\right)}{(\bar h-1)\bar h}\left(-2 \zeta _3-\frac{1}{(\bar h-1)^3\bar h^3}-\frac{2}{(\bar h-1)^2\bar h^2}+2 S_3\right)$
 \rule{0pt}{4.5ex} \rule[-4.5ex]{0pt}{0pt} \\
\end{tabular}
\caption{Inversions for $ G(\bar{z}) $ not containing explicit powers of $\log\bar z$.}\label{tab:logfree}
\end{center}
\end{table}

\begin{table}[H]
\begin{center}
\begin{tabular}{l|l}
$ G(\bar{z}) $ & $\hat{a}(J)$\rule[-1.1ex]{0pt}{0pt}  \\\hline
$\log ^2\left(1-\bar{z}\right) \log\bar z$ & $\displaystyle-\dfrac{4 \left(2 \bar{h}-1\right)}{(\bar h-1)^2\bar h^2}$
 \rule{0pt}{4.5ex} \rule[-4.5ex]{0pt}{0pt} \\
$\log ^2\left(1-\bar{z}\right)\log^2\bar z$ & $\displaystyle\dfrac{8 \left(2 \bar{h}-1\right)}{(\bar h-1)\bar h}\left(- \zeta _2+\frac{1}{(\bar h-1)^2\bar h^2}+\frac{1}{(\bar h-1)\bar h}-2 S_{-2}\right)$
 \rule{0pt}{4.5ex} \rule[-4.5ex]{0pt}{0pt} \\
$\log ^2\left(1-\bar{z}\right) \log^3\bar z$ & $\hspace{-55pt}\displaystyle\begin{matrix}\dfrac{24 \left(2 \bar{h}-1\right)}{(\bar h-1)\bar h}\left(2 \left( S_{-3}-2 S_{1,-2}\right)+ \zeta _3-\dfrac{1}{(\bar h-1)^3\bar h^3}\right.\\\qquad\qquad\qquad\qquad\qquad\left.-\dfrac{2}{(\bar h-1)^2\bar h^2}+\dfrac{ \zeta
   _2}{(\bar h-1)\bar h}+\dfrac{2 S_{-2}}{(\bar h-1)\bar h}-2 \zeta _2 S_1\right)\end{matrix}$
 \rule{0pt}{6.5ex} \rule[-6.5ex]{0pt}{0pt} \\
$\log ^3\left(1-\bar{z}\right) \log \bar z$ & $\displaystyle\dfrac{24 \left(2 \bar{h}-1\right)}{(\bar h-1)\bar h}\left(- \zeta _2+\frac{ S_1}{(\bar h-1)\bar h}-2 S_{-2}\right)$
 \rule{0pt}{4.5ex} \rule[-4.5ex]{0pt}{0pt} \\
$\log ^4\left(1-\bar{z}\right) \log\bar z$ & $\displaystyle\begin{matrix}\dfrac{48 \left(2 \bar{h}-1\right)}{(\bar h-1)\bar h}\left(-\dfrac{2S_1^2}{(\bar h-1)\bar h}-4 \left( S_{-3}-2 S_{-2,1}\right)+6 \left(S_{-3}-2
   S_{1,-2}\right)\right.\\\qquad\qquad\quad\left.+3 \zeta _3+\dfrac{2 \zeta _2}{(\bar h-1)\bar h}+\dfrac{2 S_{-2}}{(\bar h-1)\bar h}-6 \zeta _2 S_1+2 S_3\right)\end{matrix}$
 \rule{0pt}{6.5ex} \rule[-6.5ex]{0pt}{0pt} \\
$\log ^2(1-\bar{z})\text{Li}_2(1-\bar{z})\!  \log \bar{z}$ & $\hspace{-20pt}\displaystyle\begin{matrix}\dfrac{4 \left(2 \bar{h}-1\right)}{(\bar h-1)\bar h}\left(-6 \left( S_{-3}-2 S_{1,-2}\right)-\dfrac{ \zeta
   _2}{(\bar h-1)\bar h}\right.\\\qquad\qquad\qquad\qquad\qquad\left.-3 \zeta _3-\dfrac{2 S_{-2}}{(\bar h-1)\bar h}+6 \zeta _2 S_1\right)\end{matrix}$
  \rule{0pt}{6.5ex} \rule[-6.5ex]{0pt}{0pt} \\
\end{tabular}
\caption{Inversions for $ G(\bar{z}) $ containing explicit powers of $\log\bar z$.}\label{tab:logcont}
\end{center}
\end{table}

\section{Double discontinuity at fourth order}
\label{ddisc}
To order $g^4$ the terms contributing to the double discontinuity from the bilinear operators are
\begin{align}
I_{\varphi^2}=& \log^4(1-\bar z) \frac{1}{192} (\log \bar z-\log z) + \log^3(1-\bar z) \frac{1}{24} (\text{Li}_2(1-\bar z)+3\log z-3\log \bar z+2\zeta_2)\nonumber\\
&+ \log^2(1-\bar z)\Bigg(-\log z \frac{46+3 \text{Li}_2(1-\bar z)+\log\bar z+12 \zeta_2}{48} + \frac{5}{8} \text{Li}_3(1-\bar z) +\frac{1}{2} \text{Li}_3 \left(\frac{\bar z-1}{\bar z} \right)\nonumber \\
&\quad+ \frac{2 (23+6 \zeta_2) \log \bar z-\text{Li}_2(1-\bar z) (21 \log \bar z+34)-106 \zeta_2-4 \log^3\bar z+\log ^2\bar z+24 \zeta_3}{48}\Bigg),\\
I_{2} =& \frac{1}{8}\log^2(1-\bar z) \left( \log z (\zeta_2-2)  +2 \log \bar z+ \frac{1}{6}\log^3 \bar z +\text{Li}_3(1-\bar z)-\text{Li}_3\left(\frac{\bar z-1}{\bar z}\right)-\zeta_3\right).
\end{align}
In order to compute the first expression we used the value of the OPE coefficient for the bilinear scalar operator, which is fixed by crossing to be $a_0=2(1-g-g^2+\cdots)$, and also used the precise relation between $g$ and $\epsilon$.
%
%%%%%%%%

\bibliographystyle{JHEP}
\bibliography{epsilonexpansion}

\end{document}